\documentclass[aps,twocolumn,showkeys,reprint,amsmath,amssymb,prb]{revtex4-1}

\usepackage{graphicx,color}
\usepackage{dcolumn}
\usepackage{hyperref}
\usepackage{bm}
\usepackage{stmaryrd}
\usepackage{latexsym}
\usepackage{amssymb}
\usepackage{amsfonts}
\usepackage{amsmath}
\usepackage{epstopdf}

\begin{document}

\title{Two-Dimensional Superconductivity at the Titanium Sesquioxide Heterointerface}

\author{Lijie Wang,$^{\dag}$ Wenhao He,$^{\dag}$ Guangyi Huang,$^{\dag}$ Huanyi Xue,$^{\dag}$ Guanqun Zhang,$^{\dag}$ Gang Mu,$^{\ddag}$ Shiwei Wu,$^{\dag,\S}$ Zhenghua An,$^{\dag,\S}$ Changlin Zheng,$^{\dag}$ Yan Chen,$^{\dag}$ and Wei Li$^{\dag,*}$ }

\affiliation
{$^{\dag}$State Key Laboratory of Surface Physics and Department of Physics, Fudan University, Shanghai 200433, China\\
 $^{\ddag}$State Key Laboratory of Functional Materials for Informatics, Shanghai Institute of Microsystem and Information Technology, Chinese Academy of Sciences, Shanghai 200050, China\\
 $^{\S}$Institute for Nanoelectronic Devices and Quantum Computing, Fudan University, Shanghai 200433, China
 }

\date{\today}

\begin{abstract}
The study of exotic superconductivity in two dimensions has been a central theme in solid state and materials research communities. Experimentally exploring and identifying new fascinating interface superconductors with a high transition temperature ($T_c$) is challenging. Here, we report an experimental observation of intriguing two-dimensional superconductivity with a $T_c$ up to 3.8 K at the interface between Mott insulator Ti$_2$O$_3$ and polar semiconductor GaN. At the verge of superconductivity, we also observe a striking quantum metallic-like state, demonstrating that it is  a precursor to the two-dimensional superconductivity as the temperature is decreased. Our work is the first time finding this intriguing superconducting state at the heterointerface, which not only brings a new broad of perspective on the emergent quantum phenomena at the heterointerfaces but also sheds new light on exploiting the novel heterointerface superconductivity with high-$T_c$ \textit{via} heterostructure engineering.
\end{abstract}


\maketitle


\noindent Heterostructure engineering represents a state-of-the-art design strategy for realizing intriguing quantum phenomena towards next-generation quantum technologies because the symmetry breaking and the particular interactions found at the interface between two constituent materials are expected to promote novel electronic phases that are not always stable as bulk phases~\cite{Mannhart2010,Zubko2011,Hwang2012}. One of the most striking phenomena is interfacial superconductivity~\cite{Saito2017}, which has been experimentally observed at the polarized interface of two band insulators LaAlO$_3$/SrTiO$_3$~\cite{Reyren2007}, the cuprate-based interface between La$_{1.55}$Sr$_{0.45}$CuO$_4$ and La$_2$CuO$_4$~\cite{Gozar2008}, and the oxide-insulator/KTaO$_3$ interface~\cite{Liu2021}. In addition, a prominent interface-enhanced superconductivity has been reported as high as 65 K in single-layer FeSe films epitaxially grown on SrTiO$_3$ substrates compared to bulk FeSe displaying superconductivity below 8 K~\cite{Wang2012,FeSe,Hsu2008}, demonstrating that the electron correlation and interface effects cooperatively contribute to the remarkable enhancement of superconductivity~\cite{Lee2014}.

Titanium oxides are a large family of transition metal materials with many fascinating and appealing electrical properties and have potential applications as a consequence of the variable oxidation states associated with strong structure-property correlations~\cite{Jeong2012,Yang2013,ZhangC,YoshimatsuK}. Among them, titanium sesquioxide Ti$_2$O$_3$ is a typical antiferromagnetic Mott insulator with strong electron correlation and a narrow-band gap of 0.1 eV as a result of  partial filling of the Ti $3d^1$ band $(S=1/2)$~\cite{Morin1959,Nakatsugawa}, which is similar to the configuration in the Mott insulator LaTiO$_3$~\cite{Keimer2000} and indicates promise as a candidate unconventional superconductor driven by electron correlation effects upon charge carrier injection~\cite{PALee2006}. Interestingly, the superconductivity has been experimentally observed in Ti$_2$O$_3$ polymorphic films with the orthorhombic phase epitaxially grown on the $\alpha$-Al$_2$O$_3$ substrates, whereas the corundum phase of Ti$_2$O$_3$ polymorphic films remains an insulator, in recent years~\cite{Li_2018,Li2018}. The three-dimensional isotropic superconductivity in the Ti$_2$O$_3$/Al$_2$O$_3$ heterostructure has also been revealed by angular-dependent upper critical fields~\cite{Feng2020}. Notably, the charge carrier concentration is found as high as $10^{21}$ cm$^{-3}$ in the orthorhombic Ti$_2$O$_3$/Al$_2$O$_3$~\cite{Li2018}, which is approximately one order larger than that of corundum films, demonstrating that a large amount of oxygen vacancies induced electron-like charge carriers drives the Mott insulator Ti$_2$O$_3$ into a superconductivity. 

\begin{figure*}[t!]
\centering
\includegraphics[bb=85 30 345 295,width=11cm,height=10cm]{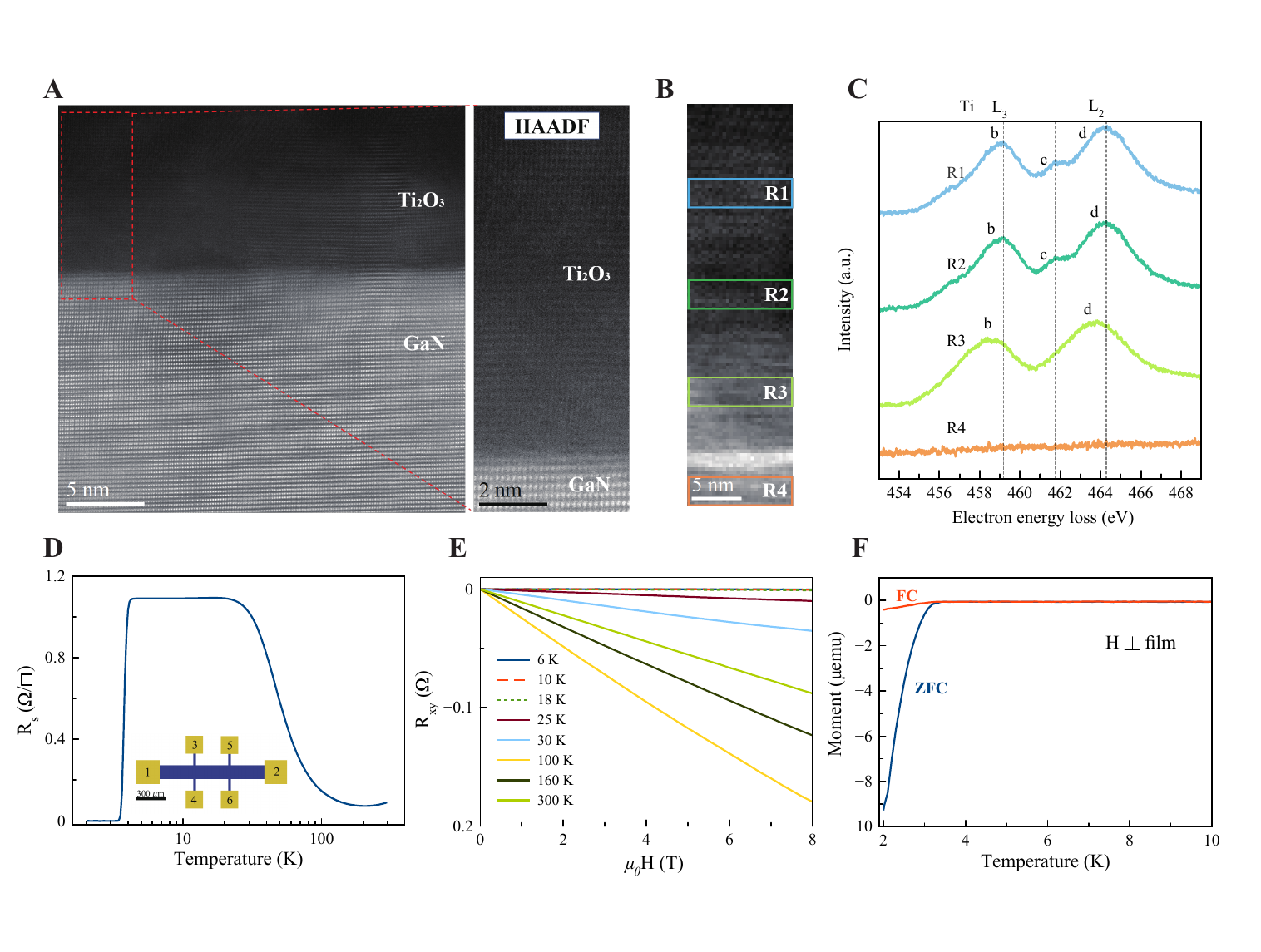}
\caption{\textbf{Structural and electronic properties of Ti$_2$O$_3$/GaN heterostructure. (A) HAADF-STEM image of Ti$_2$O$_3$/GaN. (B) Integrated EELS spectra windows around the interface of Ti$_2$O$_3$/GaN, and (C) corresponding EELS spectra. (D) Temperature dependence of the sheet resistance R$_{\mathrm{s}}$ at zero magnetic field for Ti$_2$O$_3$/GaN. The Hall bar structure is schematically illustrated in the inset of (D). (E) Transverse Hall resistance R$_{\mathrm{xy}}$ versus out-of-plane magnetic field for various fixed temperatures. (F) Temperature dependence of the DC magnetization of Ti$_2$O$_3$/GaN in FC and ZFC modes with an applied out-of-plane magnetic field of 10 Oe. The thickness of the film is 108 nm.}}
\label{fig1}
\end{figure*}

In this work, we design an experimental study on the Mott insulating Ti$_2$O$_3$ polymorphic films grown on the polar semiconductor GaN with a wide-band gap and a high piezoelectric constant~\cite{Bernardini1997}, enabling strong interfacial coupling effects that prompt intriguing quantum phases at their interface. Electrical transport and magnetization measurements on the as-grown films reveal the appearance of two-dimensional anisotropic superconductivity at a $T_c$ of 3.8 K. Furthermore, at the verge of superconductivity, we also observe a wide range of temperature-independent electrical resistance plateau associated with vanishing Hall resistance as a manifestation of quantum metallic-like state within the Bose-metal scenario. By tuning the thickness of Ti$_2$O$_3$ films, the emergence of the quantum metallic-like state accompanies the appearance of superconductivity as the temperature decreases, demonstrating that the interfacial superconductivity essentially evolves from the quantum metallic-like state. Compared to the absence of both the anisotropic superconductivity and the quantum metallic-like state in Ti$_2$O$_3$/Al$_2$O$_3$~\cite{Li2018}, we attribute these emergent quantum phases to be the intrinsic property of the Ti$_2$O$_3$/GaN heterointerface boosted by the cooperative effects of the electron correlation and the strong interfacial coupling between Ti$_2$O$_3$ films and polar GaN.

\notag\

\noindent\textbf{RESULTS AND DISCUSSION}

\noindent The Ti$_2$O$_3$ thin films are grown on a (0001)-oriented GaN substrate with an N-polar terminated face by pulsed laser deposition. X-ray diffraction measurements suggest the as-grown Ti$_2$O$_3$ film to be polycrystalline with an admixture of corundum and orthorhombic polymorphs (Figure S2), in agreement with previous studies~\cite{Li_2018,Li2018,Feng2020}. The microstructure of the interface of Ti$_2$O$_3$/GaN is also examined by aberration-corrected scanning transmission electron microscopy (STEM). From the high angle annular dark field (HAADF)-STEM image shown in Figure~\ref{fig1}A, abrupt interface between an annealed GaN surface and the polycrystalline Ti$_2$O$_3$ overlayer could be clearly resolved. Looking at the sample from a larger field of view, the thickness of the Ti$_2$O$_3$ film is found to be about 110 nm, consistent with the measurements of a profilometer. Monochromated STEM-electron energy loss spectroscopy (EELS) spectrum imaging is further utilized to clarify the electronic states of the film near the interface. The EELS signals are integrated within four different regions (Figure~\ref{fig1}B) and shown in Figure~\ref{fig1}C. Apart from the interface (R1 and R2 regions), the extra peak $c$ between the Ti L$_{3,2}$ edges displays the typical feature of Ti$^{3+}$ in Ti$_2$O$_3$ (see Figure~\ref{fig1}C), consistent with the previous study in the bulk of Ti$_2$O$_3$~\cite{ref41}. Approaching the interface (the R3 region), notably, the EELS spectrum changes dramatically that the peak $c$ is disappeared (see Figure~\ref{fig1}C). Additionally, the two main peaks of Ti L$_{3,2}$ edges shift towards low energy loss of $\sim$0.5 eV by compared with that in R1 and R2 regions. Eventually, the overall EELS peaks disappear completely in GaN substrate (the R4 region). Similar EELS features are also observed in various sample grown under the same conditions (Section I in Supporting Information). Therefore, these EELS results unambiguously demonstrate that the valence state of Ti decreases towards Ti$^{2+}$ at around the interface of Ti$_2$O$_3$/GaN~\cite{ref41}, suggestive of the existence of electron charge transfer from polar GaN substrate to the Ti$_2$O$_3$ layers, which is responsible for the emergence of two-dimensional electron gas at their interface. This mechanism is similar to the nature of the two-dimensional electron gas formed at the interface between two band insulators polar-LaAlO$_3$ and nonpolar-SrTiO$_3$~\cite{Huijben}.

Figure~\ref{fig1}D displays the sheet resistance R$_{\mathrm{s}}$ as a function of temperature for the Ti$_2$O$_3$/GaN heterostructure measured in a Hall bar structure, schematically illustrated in the inset of Figure~\ref{fig1}D. The electrical resistance R$_{\mathrm{s}}$ at a room temperature of 300 K is 0.09 $\Omega$. Such a low electrical resistance R$_{\mathrm{s}}$ implies the existence of high charge carrier concentrations and electron mobility induced by a combination of the polar nature of GaN and oxygen vacancies at the Ti$_2$O$_3$/GaN interface (see Figure~\ref{fig1}C). Through the Hall effect measurements shown in Figure~\ref{fig1}E, the transverse Hall resistance R$_{\mathrm{xy}}$ reveals that the charge carriers in the Ti$_2$O$_3$/GaN are electrons, and the carrier density is estimated to be 5.67$\times$10$^{16}$ cm$^{-2}$, approximately one order of magnitude smaller than that of orthorhombic Ti$_2$O$_3$/Al$_2$O$_3$~\cite{Li2018}. 
The electron mobility is thus evaluated to be 1208 cm$^2$V$^{-1}$s$^{-1}$ at 300 K. 
At temperatures lower than 200 K, the sample exhibits an increase in the resistance characteristic of variable range hopping-like conductivity~\cite{VRH2002} (also see Section II in Supporting Information), which then saturates at a resistance plateau with a wide temperature range of 20 K. To the best of our knowledge, this striking phenomenon in the pristine film has never been reported in the previous studies. The temperature-independent electrical resistance plateau is strongly reminiscent of a quantum metallic-like state~\cite{Kapitulnik2019,Fisher1989,Phillips2003,Saito2015,Tsen2016,Yangchao} within the hallmark of the Bose-metal scenario in the Ti$_2$O$_3$/GaN in which a two-dimensional system of interacting bosons may form a gapless, nonsuperfluid state, such as a candidate of bipolaronic state resulting from the strong interfacial effect of polar GaN~\cite{Alexandrov1995}. Remarkably, the Hall resistance R$_{\mathrm{xy}}$ is found to be zero (Figure~\ref{fig1}E) as a result of an inherent particle-hole symmetry~\cite{Breznay2017}, unambiguously providing the strong compelling evidence for the intrinsic behavior of this experimentally observed quantum metallic-like state (also see Section III in Supporting Information). 
Here, it should be noted that this quantum metallic-like state observed in Ti$_2$O$_3$/GaN at a finite temperature behaves in a quite different mechanism by compared to the anomalous metallic state with a finite value of transverse Hall resistance in cuprate~\cite{Ando2004}.
Although qualitatively similar behavior has also been previously reported in disordered ultrathin Ga films~\cite{Christiansen2002} and gated heterostructured superconductors~\cite{Biscaras2012,Chen2018,Chen2021}, the origin of such a quantum metallic state remains to be intensely debated~\cite{Kapitulnik2019,Fisher1989,Phillips2003}. As the temperature is further decreased, notably, the electrical resistance R$_{\mathrm{s}}$ undergoes a narrow and sharp transition with a width of less than 0.5 K to a zero-resistance state, measured to the limit of our instrument resolution. The critical temperature is determined to be $T_c= 3.8$ K, as defined by where the resistance is at the midpoint of the plateau value. Temperature-dependent direct current (DC) magnetization measurements in both field-cooling (FC) and zero-field-cooling (ZFC) modes under an applied out-of-plane magnetic field of 10 Oe are also carried out to further examine the nature of the zero-resistance state in Ti$_2$O$_3$/GaN, as shown in Figure~\ref{fig1}F. The observed negative magnetic susceptibility, which is the ratio of the measured magnetization to the applied magnetic field, indicates the diamagnetism induced by the Meissner effect, unambiguously confirming the appearance of superconductivity below $T_c$.

\begin{figure}[t!]
\centering
\includegraphics[bb=25 45 420 340,width=8cm,height=6.2cm]{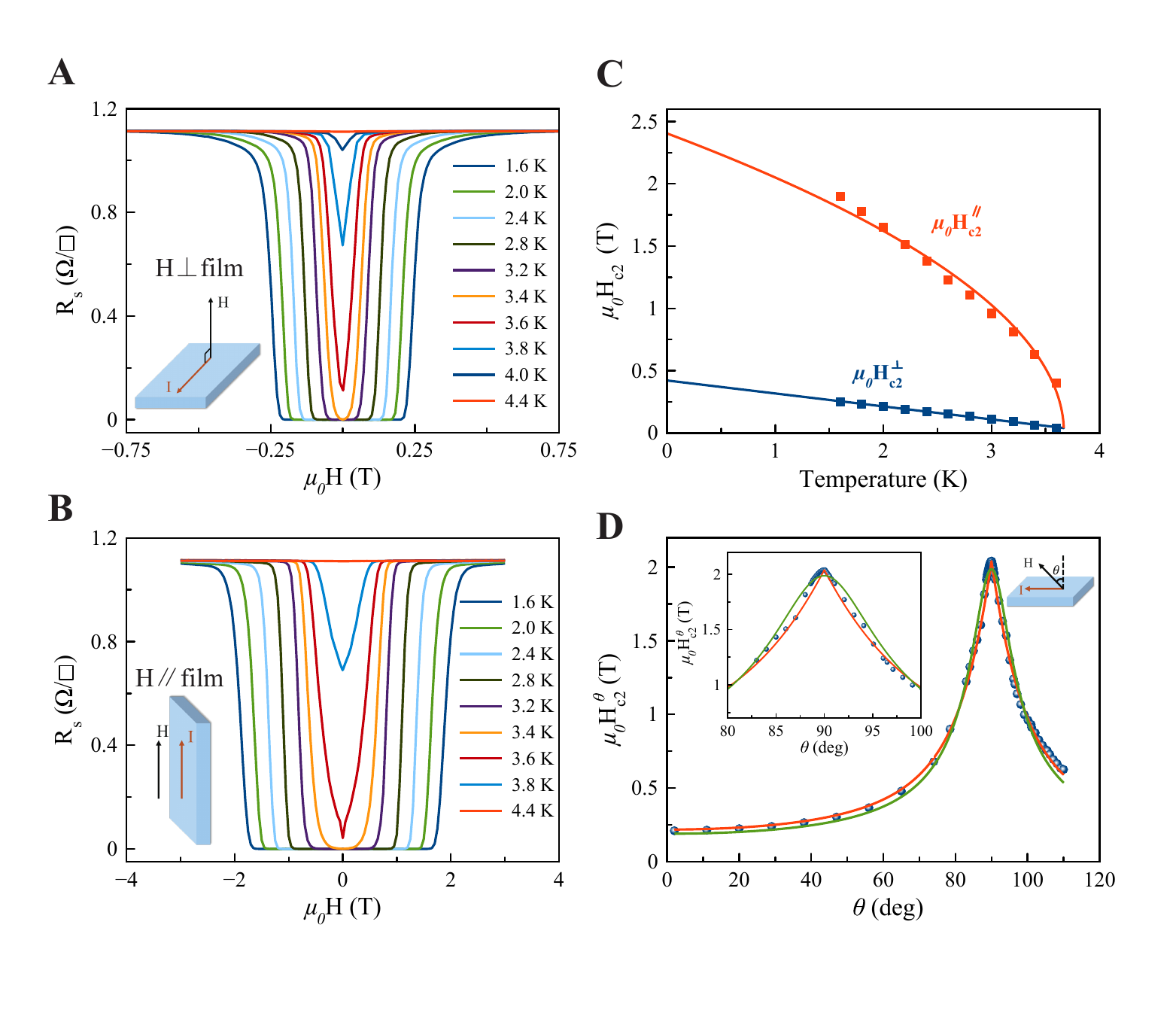}
\caption{\textbf{Anisotropic superconductivity of Ti$_2$O$_3$/GaN. Magnetoresistance for fields (A) perpendicular and (B) parallel to the sample plane surface of Ti$_2$O$_3$/GaN. (C) Temperature dependence of the upper critical field $\mu_0$H$_{c2}$, which is determined at half the value of R$_{\mathrm{s}}$ in (A) and (B). (D) Angular dependence of the upper critical fields $\mu_0$H$_{c2}^{\theta}$ ($\theta$ represents the angle between a magnetic field and the perpendicular direction to the surface of Ti$_2$O$_3$/GaN). The inset shows a close-up of the region around $\theta$ = 90$^{\circ}$. The red and green solid lines are the theoretical representations of H$_{c2}^{\theta}$, using the two-dimensional Tinkham formula and the three-dimensional anisotropic G-L model, respectively.}}
\label{fig2}
\end{figure}

\begin{figure}[t!]
\centering
\includegraphics[bb=25 20 440 285,width=8cm,height=6cm]{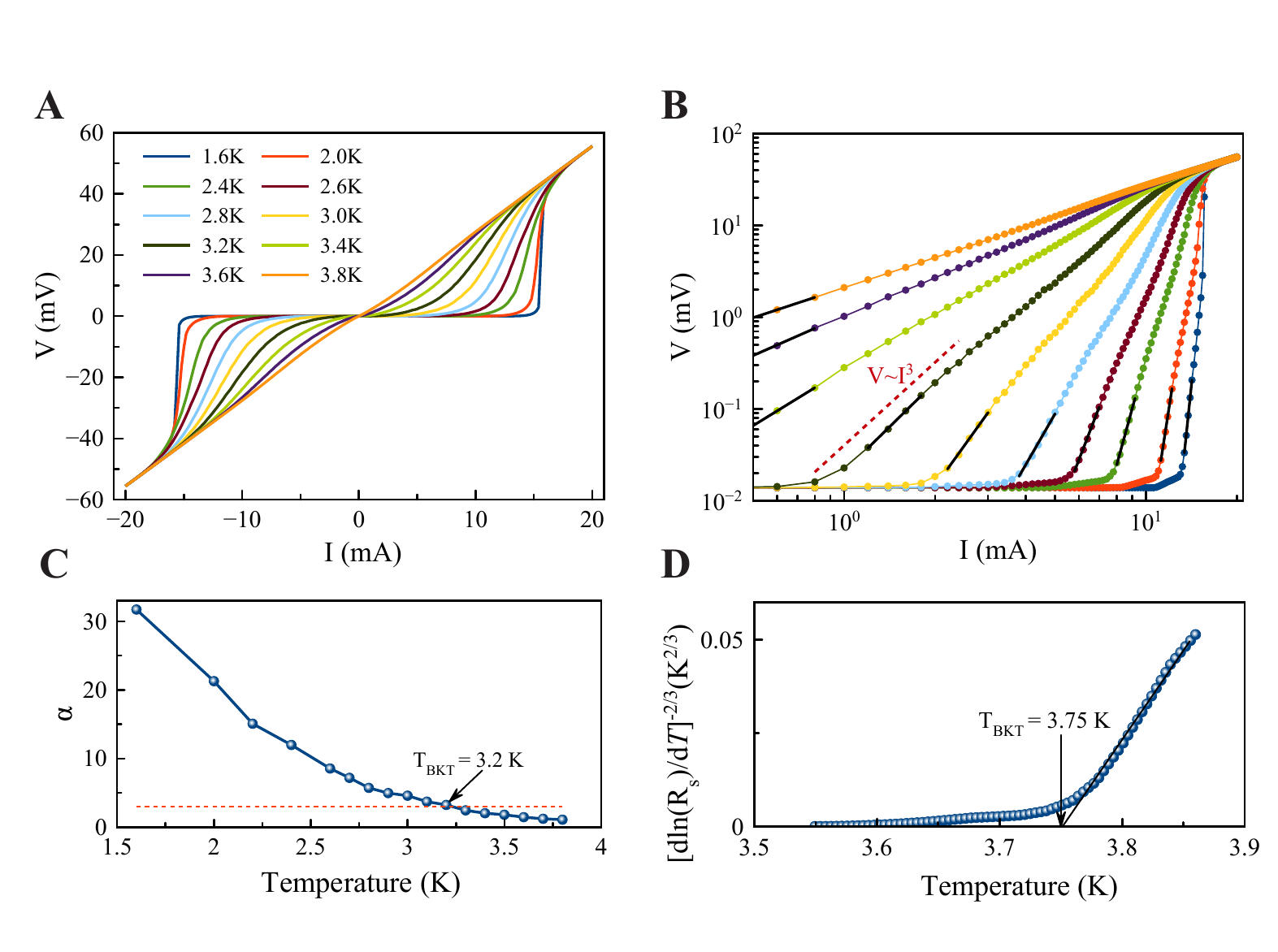}
\caption{\textbf{Two-dimensional superconducting behavior of Ti$_2$O$_3$/GaN. (A) Temperature-dependent $I$-$V$ measurements. (B) Corresponding logarithmic scale representation of (A). The long red dashed line denotes the $V\propto I^3$ dependence. (C) Temperature dependence of the power-law exponent $\alpha$, as deduced from the fits shown in (B). (D) R$_{\mathrm{s}}(T)$ dependence of the same sample, plotted on a $[d\mathrm{ln}(\mathrm{R}_{\mathrm{s}})/dT]^{-2/3}$ scale. The solid line is the behavior expected for a BKT transition with $T_{\mathrm{BKT}} = 3.75$ K.}}
\label{fig3}
\end{figure}

To further clarify the superconducting properties of Ti$_2$O$_3$/GaN heterostructures, we measure the magnetoresistance R$_{\mathrm{s}}$($\mu_0$H) (here, $\mu_0$ is the vacuum permeability) at various temperatures between 1.6 and 4.4 K with fields perpendicular ($\mu_0$H$_\perp$) and parallel ($\mu_0$H$_\parallel$) to the sample plane surface, as shown in Figure~\ref{fig2}A,B, respectively. Fundamental superconducting behaviors are clearly observed for both the superconducting upper critical fields $\mu_0$H$_{c2}^\perp$ and $\mu_0$H$_{c2}^\parallel$, which shift to a lower value with increasing temperature, where $\mu_0$H$_{c2}$ is defined as the magnetic field at the midpoint of the electrical resistance transition. Notably, the magnetoresistance R$_{\mathrm{s}}$($\mu_0$H) varies differently with $\mu_0$H$_\perp$ and $\mu_0$H$_\parallel$. For example, R$_{\mathrm{s}}$ as a function of $\mu_0$H$_\perp$ reaches the normal resistance at $\sim$0.25 T (the upper critical field $\mu_0$H$_{c2}^\perp$), which is significantly smaller than $\mu_0$H$_{c2}^\parallel= 1.9$ T for the parallel field at 1.6 K. This strong anisotropy in the observed upper critical fields provides an indication of a two-dimensional superconducting feature in Ti$_2$O$_3$/GaN heterostructures. The temperature-dependent upper critical fields $\mu_0$H$_{c2}$ derived from the magnetoresistance R$_{\mathrm{s}}$($\mu_0$H) curves in Figure~\ref{fig2}A,B are shown in Figure~\ref{fig2}C and are well fitted by the phenomenological two-dimensional Ginzburg-Landau (G-L) model~\cite{Tinkham1996}: $\mu_0$H$_{c2}^\perp(T)=\frac{\Phi_0}{2\pi\xi_{\mathrm{GL}}^2}(1-\frac{T}{T_c})$ and $\mu_0$H$_{c2}^\parallel(T)=\frac{\Phi_0\sqrt{12}}{2\pi\xi_{\mathrm{GL}}d_{\mathrm{SC}}}\sqrt{1-\frac{T}{T_c}}$,where $\Phi_0$, $\xi_{\mathrm{GL}}$, and $d_{\mathrm{SC}}$ denote a flux quantum, the in-plane superconducting coherence length at $T = 0$ K, and the effective thickness of superconductivity, respectively.
Using the extrapolated $\mu_0$H$_{c2}^\perp(0) = 0.42$ T and $\mu_0$H$_{c2}^\parallel(0) = 2.4$ T, we find $\xi_{\mathrm{GL}}= 27.8$ nm and $d_{\mathrm{SC}}=17$ nm, where the $\xi_{\mathrm{GL}}$ is substantially larger than $d_{\mathrm{SC}}$, suggestive of a two-dimensional superconductivity nature. Additionally, the detailed out-of-plane polar angle ($\theta$) dependence of the upper critical field H$_{c2}^\theta$ at a fixed temperature of 2 K is carried out to further quantitatively verify the two-dimensional behavior of the superconducting Ti$_2$O$_3$/GaN (see Figure S8A). By projecting the magnetic fields onto the out-of-plane orientation, H$_{\perp}=$H$\cos\theta$, all the data with various values of magnetoresistance R$_{\mathrm{s}}$ collapse onto a single curve (see Figure S8B), providing an additional evidence for the two-dimensional character of the superconductivity in Ti$_2$O$_3$/GaN. In Figure~\ref{fig2}D, we also show the upper critical field $\mu_0$H$_{c2}^\theta$ as a function of polar angle $\theta$ extracted from Figure S8A. The data are fitted with the two-dimensional Tinkham formula (red solid curve) and the three-dimensional anisotropic G-L model (green solid curve), given by $\frac{\mathrm{H}_{c2}^\theta \lvert \cos\theta \lvert}{\mathrm{H}_{c2}^\perp} + (\frac{\mathrm{H}_{c2}^\theta \sin\theta}{\mathrm{H}_{c2}^\parallel})^2=1$ and $(\frac{\mathrm{H}_{c2}^\theta \cos\theta}{\mathrm{H}_{c2}^\perp})^2 + (\frac{\mathrm{H}_{c2}^\theta \sin\theta}{\mathrm{H}_{c2}^\parallel})^2=1$, respectively~\cite{Tinkham1963,Lu2015,Matsuoka2020}. As expected a cusp-like peak is clearly resolved at $\theta =$ 90$^{\circ}$, which is qualitatively distinct from the three-dimensional anisotropic G-L model but is well described by the two-dimensional Tinkham model, as frequently observed in interfacial superconductivity~\cite{Reyren2007,Liu2021} and layered transition metal dichalcogenides~\cite{Lu2015,Jiang2020}.

\begin{figure*}[t!]
\centering
\includegraphics[bb=50 5 385 205,width=14cm,height=8cm]{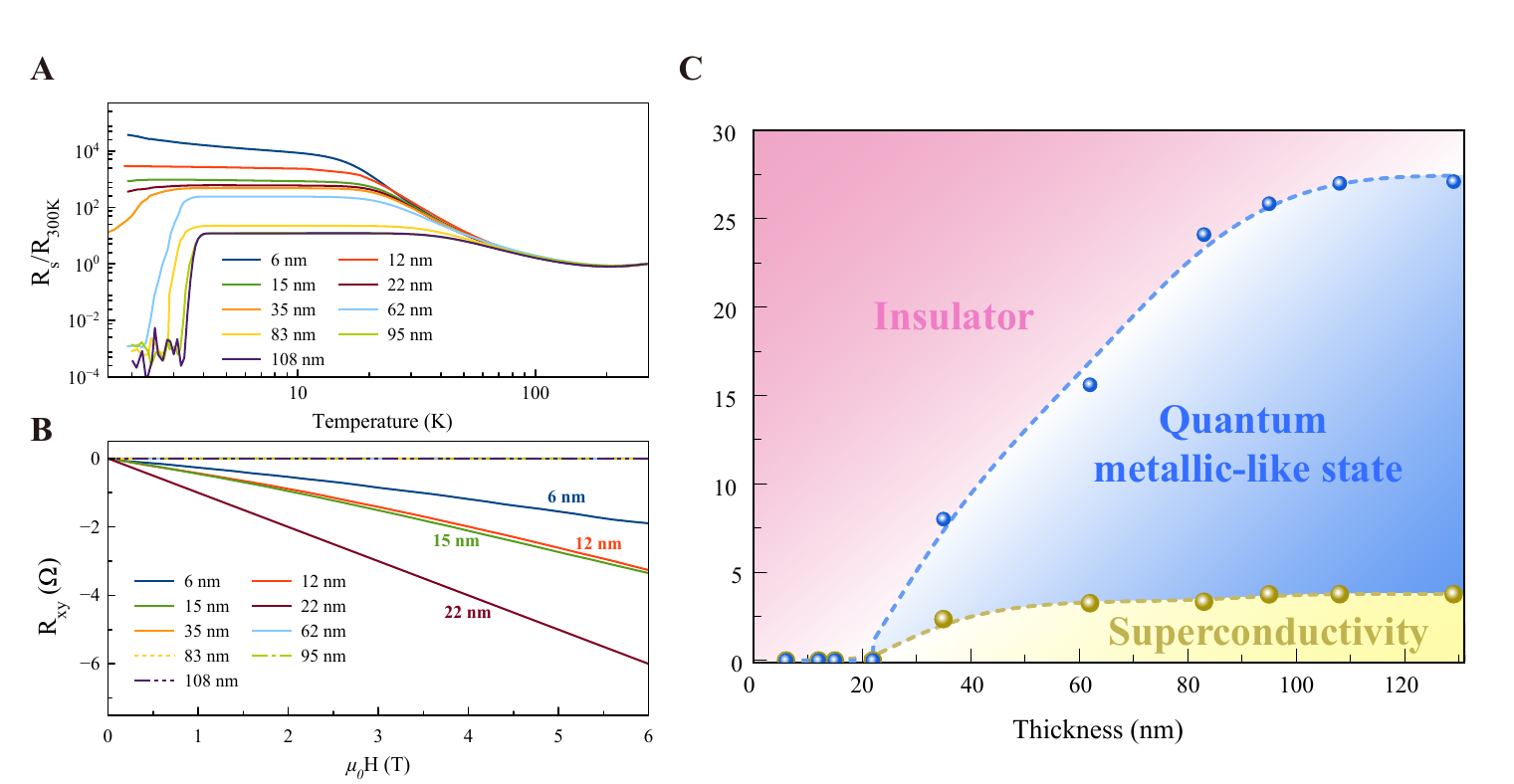}
\caption{\textbf{Thickness-dependent superconductivity in Ti$_2$O$_3$/GaN. (A) Thickness dependence of the rescaled sheet resistance R$_{\mathrm{s}}$ as a function of temperature. R$_{\mathrm{300K}}$ denotes the electrical resistance measured at a temperature of 300 K. (B) Thickness dependence of the transverse Hall resistance R$_{\mathrm{xy}}$ versus applied out-of-plane field $\mu_0$H relation at 18 K. (C) Thickness-dependent phase diagram of Ti$_2$O$_3$/GaN including the insulator, quantum metallic-like state, and superconductivity. Here, it should be noted that the quantum metallic-like state is characterized by the plateau of longitudinal resistance and simultaneously vanishing Hall resistance.}}
\label{fig4}
\end{figure*}

Since the superconductivity at the Ti$_2$O$_3$/GaN heterointerface is two-dimensional, one naturally expects fluctuations to play a crucial role in and the Berezinskii-Kosterlitz-Thouless (BKT) transition to control the establishment of phase coherence~\cite{Kosterlitz1972,Beasley1979}. In this scenario, the low-temperature, superconducting phase consists of bound vortex-antivortex pairs created by thermal fluctuations. Upon heating, the pairs dissociate and may move, inducing dissipation. The BKT temperature defines the vortex unbinding transition and can be determined using current-voltage ($I$-$V$) measurements as a function of temperature $T$, as shown in Figure~\ref{fig3}A. Below $T_c$, we find a clear critical current $I_c$, whose value decreases with increasing measurement temperature. This is a further evidence for the existence of superconductivity in Ti$_2$O$_3$/GaN heterostructures. The maximal value of $I_c$ is $\sim$15 mA at 1.6 K, which is approximately more than two orders of magnitude higher than that observed at the KTaO$_3$ interfaces~\cite{Liu2021} and more than three orders of magnitude higher than that observed at the SrTiO$_3$ interfaces~\cite{Biscaras2010,Reyren2007}. Such a significantly higher critical current value results from the high charge carrier concentration of 5.67$\times$10$^{16}$ cm$^{-2}$ at the Ti$_2$O$_3$/GaN heterointerface, which holds promise for large-scale applications in superconductor-based devices. In Figure~\ref{fig3}B, we plot $I$-$V$ on a log-log scale and observe that the slope of the $I$-$V$ characteristics smoothly evolves from the normal ohmic state, $V\propto I$, to a steeper power law resulting from the current exciting free-moving vortices, $V\propto I^{\alpha(T)}$, as superconductivity sets in at lower temperatures. At $T_{\mathrm{BKT}}$, a two-dimensional superconductor obeys the universal scaling relation $V\propto I^3$. In Figure~\ref{fig3}C, we plot $\alpha$ versus $T$, as determined by the slopes of the different $V$-$I$ traces on a log-log scale shown in Figure~\ref{fig3}B, and determine $T_{\mathrm{BKT}} = 3.2$ K, from which $\alpha = 3$ is interpolated, which is consistent with the $T_c$ as defined in Figure~\ref{fig1}D. In addition, close to $T_{\mathrm{BKT}}$, an R$_{\mathrm{s}}=$R$_0 \exp[-b(T/T_{\mathrm{BKT}}-1)^{-1/2}]$ dependence, where R$_0$ and $b$ are material parameters, is expected~\cite{Halperin1979}. As shown in Figure~\ref{fig3}D, the measured R$_{\mathrm{s}}$($T$) is also consistent with this behavior and yields $T_{\mathrm{BKT}} = 3.75$ K, in good agreement with the analysis of the $\alpha$ exponent shown in Figure~\ref{fig3}C.

To further shed light on the two-dimensional superconductivity nature and its intriguing relation to the puzzled quantum metallic-like state at the Ti$_2$O$_3$/GaN heterointerface, we carry out thickness-dependent superconductivity measurements on Ti$_2$O$_3$ films ranging from $6$ to $129$ nm in thickness measured by a profilometer. As shown in Figure~\ref{fig4}A, the thinner film of Ti$_2$O$_3$ with a thickness of $6$ nm displays insulating behavior. When the sample thickness increases, the electrical resistance decreases and saturates at the lowest temperature. If we further increase the thickness of the sample, the Ti$_2$O$_3$/GaN heterostructures undergo a transition from insulating to superconducting accompanied by the appearance of the quantum metallic-like state, indicated by a temperature-independent resistance plateau and a vanished Hall resistance, as the temperature decreases (Figure~\ref{fig4}A,B). $T_c$ increases with thickness and saturates at 3.8 K when the thickness is larger than 95 nm, as shown in Figure~\ref{fig4}C. The thickness-independent $T_c$ in the thicker films provides an additional strong evidence for the interface superconductivity at the heterointerface of Ti$_2$O$_3$/GaN. Importantly, these experimental results summarized in the phase diagram shown in Figure~\ref{fig4}C demonstrate that the quantum metallic-like state is a precursor to the interfacial superconductivity when the temperature is decreased. Here it is worth pointing out that the quantum metallic-like state observed at the Ti$_2$O$_3$/GaN heterointerfaces locates at above the $T_c$, which is experimentally reported for the first time (see Figure~\ref{fig4}C); whereas the previously reported quantum metallic phase emerges at an extreme low temperature limit, which evolves from the superconducting state \textit{via} applying an external magnetic field~\cite{Saito2015,Tsen2016} or an electric field gating~\cite{Biscaras2012,Chen2018,Chen2021}. Consistent results are also observed for all our thin films, and these results are highly reproducible, indicating that these experimental findings, including the quantum metallic-like state and the interfacial superconductivity (see Figure~\ref{fig4}C), are an intrinsic property of the Ti$_2$O$_3$/GaN heterointerface.

On the other hand, different to the three-dimensional isotropic superconductivity in the heterostructure of Ti$_2$O$_3$/Al$_2$O$_3$~\cite{Li2018}, where neither anisotropic superconductivity nor quantum metallic-like state has been observed, we thus demonstrate these emergent quantum phases observed at the heterointerface of Ti$_2$O$_3$/GaN to be the intrinsic nature of heterointerface boosted by the cooperative effects of the electron correlation and the strong interfacial coupling between Ti$_2$O$_3$ films and polar GaN. 

At last, we also address a theoretical interpretation, which should give us a hint to understand the underlying mechanism of interface superconductors. The strong interfacial coupling \textit{via} electron-phonon interaction with an optical phonon mode of polar GaN substrate will significantly modulate the electron kinetic energy at the interface of Ti$_2$O$_3$/GaN, and enhance the effective mass of electrons, manifesting a polaron-like quasiparticle~\cite{Alexandrov1995}. This quasiparticle could be stabilized by the oxygen vacancy defects and its electrical resistance increases with decreasing temperature behaved as variable range hopping-like feature. As further decreasing the temperature, polarons are close together and interact with each other, which gives rise to a strongly bound bipolaron as a precursor of preformed electron Cooper pair. Remarkably, such bipolarons preserve particle-hole symmetry and exhibit the behaviors of quantum metallic-like state with vanishing Hall resistance. Furthermore, we expect that these bipolarons could collectively condense into a macroscopically phase-coherent superconducting state eventually when the temperature is further decreased~\cite{PreformedCooper}. The physical picture of bipolaronic effects induced by the polar GaN essentially captures the overall nature of emergent quantum metallic-like state and interfacial superconductivity observed at the interface of Ti$_2$O$_3$/GaN shown in Figure~\ref{fig4}C. Therefore, we suggest that the interface superconducting Ti$_2$O$_3$/GaN is a strong candidate for bipolaronic superconductivity, which has long been a topic of interest sought in condensed matter physics and material science~\cite{Bipolaron}. Further experiments, including probing of the temperature-dependent tunneling spectra of electronic state evolution at the interface, will be helpful for elucidating the underlying bipolaronic nature of these intriguing quantum phases that we observe.

\notag\

\noindent\textbf{CONCLUSION}

\noindent In conclusion, we have experimentally observed an intriguing interfacial superconductivity at the heterointerface of Ti$_2$O$_3$/GaN, which evolves from a quantum metallic-like state with vanishing Hall resistance inherent to the particle-hole symmetry as decreasing the temperature, manifesting an appealing bipolaronic superconductivity in nature. In particular, our finding of novel interfacial superconductivity is unexpected, which will certainly stimulate a great deal of both experimental and theoretical studies on the heterostructures to exploit the emergent new fascinating interface superconductors with high-$T_c$ \textit{via} heterostructure engineering.

\notag\

\noindent\textbf{METHODS}

\textbf{Thin film growth and structural characterizations.} Ti$_2$O$_3$ thin films are grown on GaN (0001) substrates by pulsed laser deposition in an ultrahigh vacuum chamber (base pressure of $10^{-9}$ Torr). Prior to growth, the GaN substrates are ultrasonically cleaned with acetone and ethanol. During deposition, a sintered Ti$_2$O$_3$ ceramic target (Kurt J. Lesker Company Inc.) is used to grow the Ti$_2$O$_3$ films with a KrF excimer laser (Coherent 102, wavelength: $\lambda = 248$ nm). A pulse energy of 110 mJ and a repetition rate of 10 Hz are used. The Ti$_2$O$_3$ films are deposited at 750 $^{\circ}$C in a vacuum chamber to promote growth of the superconducting phase (also see the reflection high energy electron diffraction patterns shown in Figure S1). All the samples are cooled to room temperature at a constant rate of 20 $^{\circ}$C/min in vacuum after deposition. The crystalline quality of Ti$_2$O$_3$ films under ambient conditions is examined by X-ray diffraction using the BL02U2 beamline of the Shanghai Synchrotron Radiation Facility (SSRF) with an X-ray wavelength of 0.886 \AA~and by four-circle XRD (Bruker D8 Discover, Cu K$\alpha$ radiation, $\lambda$ = 1.5406 \AA) operated in high-resolution mode using a three-bounce symmetric Ge (022) crystal monochromator. The thickness of the Ti$_2$O$_3$ films is measured by using a stylus profilometer (Bruker DektakXT).

\textbf{TEM measurements.} STEM and high-resolution EELS are performed to examine the thin film morphology and the electronic states of Ti 3$d$ near the interface of Ti$_2$O$_3$/GaN heterostructure. The TEM lamella is prepared by focused ion beam (Helios-G4-CX, Thermo Fisher Scientific) using lift-out method. The cross-section view of the sample is characterized by a field-emission transmission electron microscope (Themis Z, Thermo Fisher Scientific) fitted with double aberration correctors (SCORR and CETCOR, CEOS GmbH) and a monochromator (Thermo Fisher Scientific). The microscope is operated at 300 kV. The high-resolution EELS experiment is performed in Monochromated mode with an energy resolution of 0.25 eV. The EELS spectra are then collected using an Gatan Continuum HR/1066 energy filter system.

\textbf{Magnetization and electrical transport measurements.} Before electrical transport measurements, the magnetic properties of the Ti$_2$O$_3$ films are measured using a superconducting quantum interference device (SQUID) magnetometer (MPMS, Quantum Design Inc.). For measurement of the DC magnetization as a function of temperature, the samples are first cooled to 2 K in zero field, and then, an out-of-plane magnetic field of 10 Oe is applied. The magnetization data are collected during warming from 2 to 15 K (ZFC). In the same fixed field, the samples are then cooled to 2 K again, and the magnetization data are recollected during warming from 2 to 15 K (FC). The electrical transport measurements are performed using a cryostat (Oxford Instruments TeslatronPT cryostat system). The Hall bar structure, schematically illustrated in the inset of Figure~\ref{fig1}D, is fabricated by ion-beam etching to measure the electrical transport properties. Using a commercially available measurement apparatus, the samples are mounted on a mechanical rotator in a $^4$He cryostat to study the out-of-plane polar angle $\theta$-dependent magnetoresistance. The misalignment of the field with the film plane is estimated to be less than 7$^{\circ}$ as our experimental error.

\notag\

\noindent\textbf{Supporting Information}

\noindent
The Supporting Information includes the extra data of large scale STEM and EELS analysis of the Ti$_2$O$_3$/GaN heterointerface, and the theoretical fitting of variable-range hopping-like model for R-$T$ curve in the normal state, as well as current and magnetic field independence of quantum metallic-like state at the Ti$_2$O$_3$/GaN heterointerface.

\notag\

\noindent\textbf{AUTHOR INFORMATION}

\noindent\textbf{Corresponding Author}

\textbf{Wei Li} --- {\it State Key Laboratory of Surface Physics and Department of Physics, Fudan University, Shanghai 200433, China};
\noindent Email: w$\_$li@fudan.edu.cn

\notag\

\noindent\textbf{Authors}

\textbf{Lijie Wang} --- {\it State Key Laboratory of Surface Physics and Department of Physics, Fudan University, Shanghai 200433, China}

\textbf{Wenhao He} --- {\it State Key Laboratory of Surface Physics and Department of Physics, Fudan University, Shanghai 200433, China}

\textbf{Guangyi Huang} --- {\it State Key Laboratory of Surface Physics and Department of Physics, Fudan University, Shanghai 200433, China}

\textbf{Huanyi Xue} --- {\it State Key Laboratory of Surface Physics and Department of Physics, Fudan University, Shanghai 200433, China}

\textbf{Guanqun Zhang} --- {\it State Key Laboratory of Surface Physics and Department of Physics, Fudan University, Shanghai 200433, China}

\textbf{Gang Mu} --- {\it State Key Laboratory of Functional Materials for Informatics, Shanghai Institute of Microsystem and Information Technology, Chinese Academy of Sciences, Shanghai 200050, China}

\textbf{Shiwei Wu} --- {\it State Key Laboratory of Surface Physics, Department of Physics, and Institute for Nanoelectronic Devices and Quantum Computing, Fudan University, Shanghai 200433, China}

\textbf{Zhenghua An} --- {\it State Key Laboratory of Surface Physics, Department of Physics, and Institute for Nanoelectronic Devices and Quantum Computing, Fudan University, Shanghai 200433, China}

\textbf{Changlin Zheng} --- {\it State Key Laboratory of Surface Physics and Department of Physics, Fudan University, Shanghai 200433, China}

\textbf{Yan Chen} --- {\it State Key Laboratory of Surface Physics and Department of Physics, Fudan University, Shanghai 200433, China}

\notag\

\noindent\textbf{Author Contributions}

\noindent
W.L. conceived the project and designed the experiments. L.W. grew the samples and performed the measurements. W.H., G.H., and C.Z. performed scanning transmission electron microscopy measurements. H.X. and Z.A. fabricated the Hall bar structure on the thin films. W.L. wrote the paper. All authors discussed the results and gave approval to the final version of the manuscript.

\notag\

\noindent\textbf{Notes}

\noindent
The authors declare no competing financial interest.

\notag\

\noindent\textbf{Acknowledgements}

\noindent
This work is supported by the National Natural Science Foundation of China (Grant Nos. 61871134, 11927807, and 62171136) and Shanghai Science and Technology Committee (Grant Nos. 18JC1420400 and 20DZ1100604). The authors also thank beamline BL02U2 of the Shanghai Synchrotron Radiation Facility.

\end{document}